\documentclass[aps,prl,reprint,superscriptaddress]{revtex4-2}
\usepackage{physics}
\usepackage{mathrsfs}
\usepackage{amsmath}
\usepackage{amssymb}
\usepackage{mathtools}
\usepackage{graphicx}
\usepackage{dsfont}
\usepackage{array}

\usepackage{dcolumn}
\usepackage{bm}
\usepackage[T1]{fontenc}
\usepackage{etoolbox}

\usepackage{hyperref}
\hypersetup{
  colorlinks=true,
  linkcolor=blue,
  filecolor=blue,
  urlcolor=blue,
  citecolor=blue,
}

\begin{document}

\newcommand{\lbnlafil}{Applied Mathematics and Computational Research Division, Lawrence Berkeley National Laboratory, Berkeley, California 94720, USA}
\newcommand{\foundryafil}{Molecular Foundry Division, Lawrence Berkeley National Laboratory, Berkeley, California 94720, USA}
\newcommand{\lbnlmatafil}{Materials Science Division, Lawrence Berkeley National Laboratory, Berkeley, California 94720, USA}
\newcommand{\qnlafil}{Quantum Nanoelectronics Laboratory, Department of Physics, University of California, Berkeley, California 94720, USA}
\newcommand{\berkeleyphysafil}{Department of Physics, University of California, Berkeley, California 94720, USA}
\newcommand{\KavliENSI}{Kavli Energy NanoScience Institute at the University of California, Berkeley and the Lawrence Berkeley National Laboratory, Berkeley, California 94720, USA}
\newcommand{\ufl}{Department of Physics, University of Florida, Gainesville, Florida 32611, USA}

\newcommand{\nimsa}{Research Center for Materials Nanoarchitectonics, National Institute for Materials Science, 1-1 Namiki, Tsukuba 305-0044, Japan}

\newcommand{\nimsb}{Research Center for Electronic and Optical Materials, National Institute for Materials Science; 1-1 Namiki, Tsukuba 305-0044, Japan}

\title{Contactless cavity sensing of superfluid stiffness in atomically thin 4Hb-TaS$_2$}
\author{Trevor Chistolini}
\thanks{These authors contributed equally}
\affiliation{\berkeleyphysafil}
\affiliation{\lbnlafil}

\author{Ha-Leem Kim}
\thanks{These authors contributed equally}
\affiliation{\berkeleyphysafil}
\affiliation{\lbnlmatafil}
\affiliation{\KavliENSI}

\author{Qiyu Wang}
\thanks{These authors contributed equally}
\affiliation{\berkeleyphysafil}

\author{Su-Di Chen}
\thanks{These authors contributed equally}
\thanks{Corresponding author}
\email{sudichen@ufl.edu}
\affiliation{\berkeleyphysafil}
\affiliation{\lbnlmatafil}
\affiliation{\KavliENSI}
\affiliation{\ufl}

\author{Luke Pritchard Cairns}
\affiliation{\berkeleyphysafil}

\author{Ryan Patrick Day}
\affiliation{\berkeleyphysafil}

\author{Collin Sanborn}
\affiliation{\berkeleyphysafil}

\author{Hyunseong Kim}
\affiliation{\berkeleyphysafil}

\author{Zahra Pedramrazi}
\affiliation{\lbnlafil}

\author{Ruishi Qi}
\affiliation{\berkeleyphysafil}
\affiliation{\lbnlmatafil}
\affiliation{\KavliENSI}

\author{Takashi Taniguchi}
\affiliation{\nimsa}

\author{Kenji Watanabe}
\affiliation{\nimsb}

\author{James G. Analytis}
\affiliation{\berkeleyphysafil}
\affiliation{\lbnlmatafil}

\author{David I. Santiago}
\affiliation{\lbnlafil}

\author{Irfan Siddiqi}
\affiliation{\berkeleyphysafil}
\affiliation{\lbnlafil}
\affiliation{\KavliENSI}

\author{Feng Wang}
\thanks{Corresponding author}
\email{fengwang76@berkeley.edu}
\affiliation{\berkeleyphysafil}
\affiliation{\lbnlmatafil}
\affiliation{\KavliENSI}

\date{\today}

\begin{abstract}
The exceptional tunability of two-dimensional van der Waals materials offers unique opportunities for exploring novel superconducting phases. However, in such systems, the measurement of superfluid phase stiffness, a fundamental property of a superconductor, is challenging because of the mesoscopic sample size. Here, we introduce a contact-free technique for probing the electrodynamic response, and thereby the phase stiffness, of atomically thin superconductors using on-chip superconducting microwave resonators. We demonstrate this technique on 4Hb-TaS$_2$, a van der Waals superconductor whose gap structure under broken mirror symmetry is under debate. In our cleanest few-layer device, we observe a superconducting critical temperature comparable to that of the bulk. The temperature evolution of the phase stiffness features nodeless behavior in the presence of broken mirror symmetry, inconsistent with the scenario of nodal surface superconductivity. With minimal fabrication requirements, our technique enables microwave measurements across wide ranges of two-dimensional superconductors.
\end{abstract}

\maketitle

Superfluid phase stiffness, the free-energy cost associated with creating a spatial variation in the phase of a complex superconducting order parameter, is a key thermodynamic property of superconductors. It plays a pivotal role in determining a material’s ability to carry dissipationless currents~\cite{tinkham2004} and, in some cases, directly constrains the critical temperature~\cite{Emery1995,chen_unconventional_2022}. Moreover, its temperature dependence provides crucial information about the superconducting gap structure, offering an essential window into the pairing mechanism in unconventional systems~\cite{hardy_precision_1993,sonier_musr_2000}. While techniques for measuring the phase stiffness in bulk materials and large-area thin films are well established~\cite{klein_microwave_1993,sonier_musr_2000,he_high-precision_2016}, such measurements remain challenging in micron-scale, atomically thin van der Waals (vdW) heterostructures, an increasingly important platform for exploring novel superconducting phases~\cite{cao_unconventional_2018,zhao_time-reversal_2023}.

An emerging approach to address this challenge is to harness the sensing capabilities of on-chip superconducting resonators. These devices are widely utilized in quantum science for the detection of qubit states~\cite{blais_circuit_2021}, single photons~\cite{Mazin_MKID_rev_2009,zmuidzinas_superconducting_2012}, and material losses~\cite{murray_material_2021}. Recently, they have been employed to probe the microwave electrodynamic response in various superconductors~\cite{annunziata_tunable_2010,singh_competition_2018,phan_detecting_2022,bottcher_circuit_2024,kreidel_measuring_2024,tanaka_superfluid_2025,banerjee_superfluid_2025, jin_exploring_2025}, including superconducting two-dimensional (2D) materials~\cite{kreidel_measuring_2024,tanaka_superfluid_2025,banerjee_superfluid_2025}. At microwave frequencies well below the superconducting gap, the imaginary part of the sample's complex conductivity provides a direct measure of the superfluid phase stiffness~\cite{dressel_electrodynamics_2002}. In all previous superfluid stiffness measurements using on-chip resonators, the superconductor of interest is incorporated into the resonator via galvanic contacts. While this approach has proven effective for the materials studied, it introduces fabrication complexity and necessitates careful considerations of the contact properties. For resistive contacts, the contact resistance negatively affects the measurement of the real part of conductivity. In certain 2D materials, achieving the necessary low values of contact resistance may also be difficult. In the case of superconducting contacts, the proximity effect may influence the properties of the sample. To circumvent these obstacles, we present a contactless measurement technique that offers a simplified fabrication process and enables access to both the imaginary and real parts of conductivity.

We demonstrate our technique on exfoliated few-layer 4Hb-TaS$_2$, a vdW material composed of alternating stacks of 1T-TaS$_2$ and 1H-TaS$_2$ layers~\cite{di_salvo_preparation_1973}. This compound has attracted significant interest due to reports of unconventional properties in bulk crystals~\cite{ribak_chiral_2020,persky_magnetic_2022}, cleaved surfaces~\cite{nayak_evidence_2021}, and mesoscopic devices~\cite{almoalem_observation_2024}. In bulk crystals, specific heat and thermal conductivity measurements indicate a nodeless superconducting gap coexisting with residual gapless states~\cite{ribak_chiral_2020, wang_evidence_2025}; a nodeless behavior has also been observed in penetration depth measurements~\cite{zhou_nodeless_2025}. On the other hand, it has been proposed that the broken mirror symmetry at a 1H-TaS$_2$-terminated surface may give rise to inter-orbital pairing and point nodes in the superconducting gap~\cite{nayak_evidence_2021}. As we will show below, our temperature-dependent measurements of the superfluid phase stiffness are inconsistent with this proposed gap structure.


\begin{figure*}
  \includegraphics[]{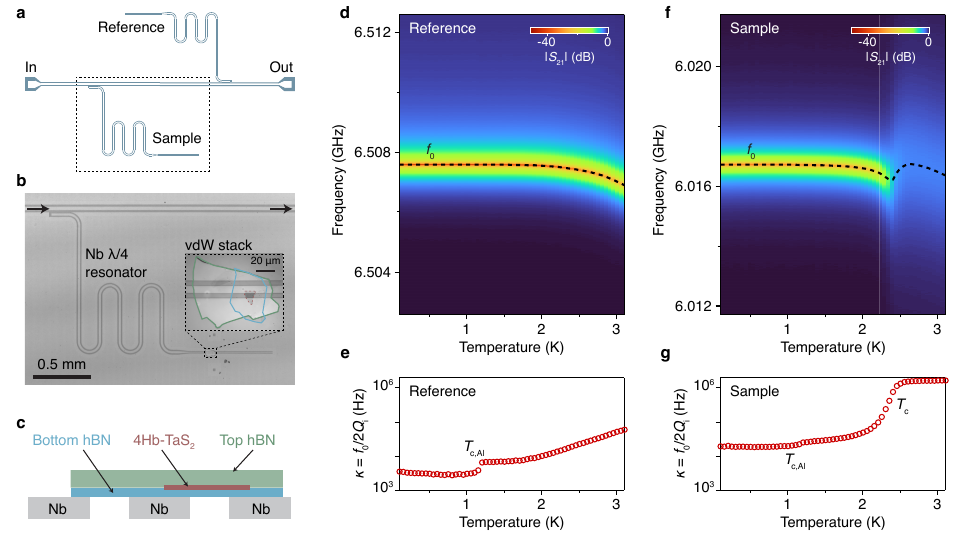}
  \caption{Device structure and microwave measurements. (a) Schematic of the chip design, showing two Nb $\lambda/4$ CPW resonators (sample and reference) of different lengths, coupled to a central transmission line in a hanger geometry. (b) Optical micrograph of the region outlined by the dashed box in (a). The vdW sample is placed away from the voltage node and bridges the signal trace and ground plane (inset). The dashed line marks the 4Hb-TaS$_2$ flake, while the solid green and blue lines denote the edges of the top and bottom hBN flakes, respectively. (c) Cross-sectional schematic of the sample region. A 4-layer 4Hb-TaS$_2$ flake is capacitively coupled to the resonator through a thin (2.2~nm) hBN dielectric. A top hBN layer serves both as encapsulation and mechanical support, whose thickness exceeds 200~nm but remains much smaller than the width of the CPW gap. (d) Amplitude of the transmission scattering parameter, $|S_{21}|$, versus frequency and temperature near the reference resonator's fundamental mode. For each temperature, the amplitude is normalized such that its maximum value away from the resonance equals unity. The dashed line shows the resonance frequency $f_0$ extracted from fits to the complex $S_{21}$ for each temperature. (e) Internal linewidth of the resonance, $\kappa \coloneq f_0/2Q_i$, as a function of temperature. (f, g), same as (d, e) for the sample resonator. $T_\text{c}$ and $T_{\text{c,Al}}$ mark the superconducting critical temperatures of the sample and the Al wirebonds connecting the CPW ground planes (not shown in (b)), respectively.}
  \label{Fig_1}
\end{figure*}

Fig.~\ref{Fig_1}(a--c) shows our device structure. Two quarter‑wavelength ($\lambda/4$) coplanar waveguide (CPW) resonators, labeled sample and reference, are patterned from a 200-nm thick Nb film on a high‑resistivity silicon substrate. While similar in design, the resonators have slightly different lengths and thus resonance frequencies. Both resonators are coupled to a common feedline in a hanger geometry, enabling characterization via measurements of the scattering parameter $S_{21}$. A 4-layer (1-unit-cell) 4Hb‑TaS$_2$ flake, exfoliated from bulk crystals and encapsulated in hexagonal boron nitride (hBN) inside a nitrogen glovebox, is transferred onto the sample resonator to a designated position away from the voltage node, such that the 4Hb‑TaS$_2$ spans over the CPW gap between the ground plane and the signal trace, similar to the capacitive-coupling design in our waveguide-based terahertz spectroscopy experiments~\cite{gallagher_quantum-critical_2019,chen_direct_2025,chen_terahertz_2025}. Further details on the device fabrication and microwave measurements are presented in the Supplemental Material~\cite{supplemental}.

In Fig.~\ref{Fig_1}(d), we plot $|S_{21}|$ for the reference resonator as a function of frequency ($f$) and temperature ($T$), which features a sharp $T$-dependent resonance. We extract the resonance frequency $f_0$ and internal linewidth $\kappa \coloneq f_0/2Q_i$ by fitting the complex $S_{21}$ spectra~\cite{supplemental,carter_scraps_2017}. Here, $Q_i$ denotes the internal quality factor of the resonator, which reaches around $10^6$ at base temperature. With increasing $T$, $f_0$ redshifts, and $\kappa$ increases (Fig.~\ref{Fig_1}(e)). A weak step in $\kappa$ is also observed, which we attribute to the superconducting transition of the Al wirebonds used to connect the CPW ground planes.

In contrast, the sample resonator exhibits markedly different behaviors. As shown in Fig.~\ref{Fig_1}(f), $f_0$ evolves non-monotonically with temperature. Moreover, $\kappa$ is around 5 times higher than that of the reference at base temperature, and with increasing $T$, rises sharply near $2.5~\text{K}$, signaling the superconductor-to-metal transition of the 4Hb‑TaS$_2$ flake (Fig.~\ref{Fig_1}(g)).

The transition temperature $T_\text{c}\sim 2.5~\text{K}$ observed here is close to the bulk $T_\text{c}$ of $2.7~\text{K}$~\cite{ribak_chiral_2020}. It is also the highest among 9 few-layer devices we fabricated for microwave or DC transport measurements~\cite{supplemental}. In all other devices, $T_\text{c}$ is significantly suppressed relative to the bulk, with no systematic dependence on thickness. This variation could arise from inhomogeneities in the bulk crystals or, more likely, the air sensitivity of exfoliated 4Hb‑TaS$_2$: the atomically thin flakes may partially degrade in the imperfect and time-varying glovebox environment prior to encapsulation. In the following, we focus on the highest-$T_\text{c}$ device, which is expected to best reflect the intrinsic properties of the material.


\begin{figure}
    \includegraphics[width=\columnwidth]{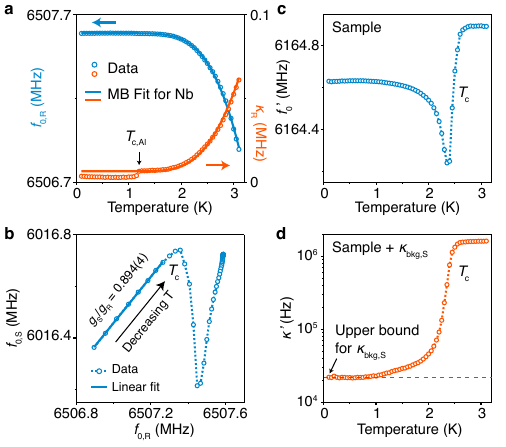}
    \caption{Separating sample response from background. (a) Cavity perturbation fit to the temperature evolutions of $f_0$ (blue) and $\kappa$ (orange) in the reference resonator, where the surface impedance of Nb is modeled using Mattis-Bardeen (MB) theory. Data at $T<T_{c,\text{Al}}$ are excluded from the fitting because the Al wirebonds are not included in the model. (b) $f_0$ of the sample resonator plotted against that of the reference at the same temperature. The ratio $g_{\text{S}}/g_{\text{R}}$ is extracted using a linear fit to the data above $T_\text{c}$. (c) Temperature evolution of the modified $f_0'$ after background subtraction. (d) Same as (c) for $\kappa'$. Dashed line indicates the upper bound of $\kappa_{\text{bkg,S}}$ from background losses unrelated to 4Hb-TaS$_2$. $T_\text{c}$ marks the superconducting transition temperature.}
    \label{Fig_2}
\end{figure}

To extract the response of the flake, we first separate its contributions to $f_0$ and $\kappa$ from those of the background by comparing the sample and reference resonators. 
The temperature dependence in the reference resonator can be modeled using cavity perturbation theory~\cite{klein_microwave_1993,dressel_electrodynamics_2002}:
\begin{equation}
\begin{split}
f_{0,\text{R}}(T) = f_{00,\text{R}} + g_{\text{R}}\cdot\text{Im}Z_{\text{Nb}}(T),\\
\kappa_{\text{R}}(T) = \kappa_{\text{bkg,R}} + g_{\text{R}}\cdot\text{Re}Z_{\text{Nb}}(T).
\label{eq_ref}
\end{split}
\end{equation}
Here, the subscript R stands for the reference resonator, $f_{00}$ represents the resonance frequency of an otherwise identical resonator with zero surface impedance, $g$ is a real constant determined by the detailed resonator geometry, $Z_{\text{Nb}}$ indicates the surface impedance of Nb, and $\kappa_{\text{bkg}}$ symbolizes a background linewidth from losses unrelated to Nb. The surface impedance of Nb is used here because the film thickness is much larger than its superconducting penetration depth~\cite{maxfield_superconducting_1965}. We also ignore the frequency dependence in $Z_{\text{Nb}}$ because the relevant frequency window is narrow~\cite{supplemental}. For the sample resonator (indicated by subscript S), we further introduce $\delta f_0$ and $\delta \kappa$ to represent the changes in $f_0$ and $\kappa$ caused by the 4Hb-TaS$_2$ flake, respectively:
\begin{equation}
\begin{split}
f_{0,\text{S}}(T) = f_{00,\text{S}} + g_{\text{S}}\cdot\text{Im}Z_{\text{Nb}}(T) + \delta f_0(T),\\
\kappa_{\text{S}}(T) = \kappa_{\text{bkg,S}} + g_{\text{S}}\cdot\text{Re}Z_{\text{Nb}}(T)+\delta \kappa(T).
\label{eq_sample}
\end{split}
\end{equation}

We fit the experimentally measured $f_{0,\text{R}}(T)$ and $\kappa_{\text{R}}(T)$ using Eqs.~\ref{eq_ref}, where $f_{00,\text{R}}$ and $\kappa_{\text{bkg,R}}$ are assumed to be constant, and $Z_{\text{Nb}}$ is calculated from Mattis-Bardeen theory~\cite{mattis_theory_1958} with a minimal set of fitting parameters as input~\cite{supplemental}. The procedure is similar to that used in our earlier work for Al airbridges~\cite{chistolini_performance_2024}. The fitting results plotted in Fig.~\ref{Fig_2}(a) well reproduce the data. We obtain a $T_\text{c}$ of 9~K and a $2\Delta_0/k_\text{B} T_\text{c}$ around 3.9 for Nb, where $\Delta_0$ denotes the zero-temperature gap size and $k_\text{B}$ is the Boltzmann constant, in good agreement with literature values~\cite{townsend_investigation_1962}.

Strictly speaking, background losses like the two-level-system (TLS) loss will lead to temperature-dependent $f_{00,\text{R}}$ and $\kappa_{\text{bkg,R}}$~\cite{Pappas_TLS}. However, the consistency between the data and the fit, as well as between the extracted Nb parameters and literature values, suggests that the temperature dependence of $f_{00,\text{R}}$ and $\kappa_{\text{bkg,R}}$ is negligible relative to the Nb-induced changes. This is further confirmed by our measurements with varying microwave power~\cite{supplemental}, where the data well overlap with each other despite the power-dependent nature of TLS loss. Similarly, for the sample resonator, because the flake-induced changes are comparable to if not larger than those from Nb, we also treat $f_{00,\text{S}}$ and $\kappa_{\text{bkg,S}}$ as temperature independent.

Having obtained $Z_{\text{Nb}}$, we still need the ratio $g_{\text{S}}/g_{\text{R}}$ to subtract out the Nb term in Eqs.~\ref{eq_sample}. As shown in Fig.~\ref{Fig_2}(b), we observe a temperature regime above the $T_\text{c}$ of 4Hb-TaS$_2$ where $f_{0,\text{S}}$ depends linearly on $f_{0,\text{R}}$. In this regime, the conductivity of 4Hb-TaS$_2$ is expected to be purely resistive and slow varying. The frequency shift thus predominantly originates from the temperature dependence of $\text{Im}Z_\text{Nb}$, and the linear slope equals $g_{\text{S}}/g_{\text{R}}$.

Using the $f_{00,\text{R}}$, $\kappa_{\text{bkg,R}}$, and $g_{\text{S}}/g_{\text{R}}$ extracted from the fits, we obtain the modified resonance frequency $f_0'$ and linewidth $\kappa'$, in which all temperature dependence originates from the 4Hb-TaS$_2$ sample and the finite-surface-impedance effects of Nb are removed:
\begin{equation}
\begin{split}
f_0' &\coloneq f_{00,\text{S}} + \delta f_0 (T) = f_{0,\text{S}} - (f_{0,R}-f_{00,\text{R}})\cdot g_{\text{S}}/g_{\text{R}},\\
\kappa' &\coloneq \kappa_{\text{bkg,S}} + \delta \kappa (T) = \kappa_{\text{S}} - (\kappa_{R}-\kappa_{\text{bkg,R}})\cdot g_{\text{S}}/g_{\text{R}}.
\end{split}
\label{eq_subtract}
\end{equation}
Although the step feature in $\kappa$ attributed to the Al wirebonds is not modeled here, its effect is minimal in comparison with the dissipation in the sample to begin with, and largely cancels out in the subtraction procedure described by Eqs.~\ref{eq_subtract}.

We plot $f_0'$ and $\kappa'$ in Fig.~\ref{Fig_2}(c) and (d), respectively. In the low-temperature regime, both $f_0'$ and $\kappa'$ are flat. As $T$ approaches $T_\text{c}$, $f_0'$ decreases sharply, then rises to a higher value before leveling off for $T>T_{\text{c}}$. In contrast, $\kappa'$ increases monotonically and exhibits a distinct step-like feature near $T_c$. These behaviors can be qualitatively understood in a cavity perturbation picture (see Supplemental Material~\cite{supplemental}). To quantitatively extract the sample conductance $\sigma = \sigma_1 + i\sigma_2$ from $f_0'$ and $\kappa'$, we employ the circuit model detailed in the End Matter.


\begin{figure}
    \includegraphics[]{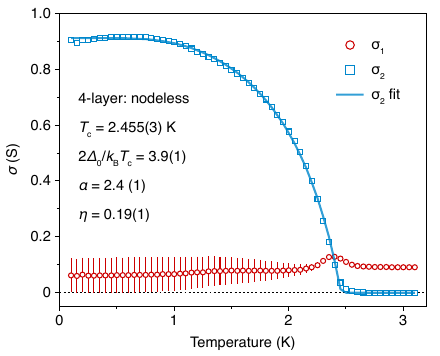}
    \caption{Temperature evolution of sheet conductance $\sigma$ at 6 GHz for the 4-layer 4Hb-TaS$_2$ flake. Data points represent the average between two limiting cases, in which $\kappa'(0.1~\text{K})$ is attributed entirely to dissipation in the sample or to background dissipation unrelated to the sample, respectively. Error bars indicate the range of $\sigma$ bounded by these two limits; those smaller than the sizes of the circle and square markers are omitted. The imaginary part, $\sigma_2$ (blue squares), is fit using the Mattis-Bardeen theory in the dirty limit with a BCS-like gap evolution (Eqs.~\ref{eq:MB2} and \ref{eq:gap}). The fit is shown in solid blue along with all fitting parameters extracted.}
    \label{Fig_Result}
\end{figure}

Fig.~\ref{Fig_Result} shows the extracted values of $\sigma_1$ and $\sigma_2$ as functions of temperature from our 4-layer 4Hb-TaS$_2$ device. The superfluid phase stiffness, $T_\theta$, can be directly obtained from $\sigma_2$ via $k_{\text{B}}T_\theta = \hbar^2\omega \sigma_2/e^2$~\cite{dressel_electrodynamics_2002}, where $\hbar$ is the reduced Planck constant and $e$ is the elementary charge. In the low-temperature limit, we find $T_{\theta} \approx 1.1\times 10^3~\text{K}$, much higher than the $T_{\text{c}}$ of the flake. Therefore, despite the sample's 2D nature, $T_\text{c}$ is predominantly set by pairing strength rather than phase stiffness. Furthermore, the temperature evolution of $T_{\theta}$ and $\sigma_2$ should only deviate from mean-field behavior within an extremely narrow window near $T_\text{c}$, which we thus ignore in the following analysis.

With increasing temperature, $\sigma_2$, and hence $T_\theta$, remain approximately constant up to $\sim 1~\text{K}$. A very weak maximum exists near 0.5~K, which is likely an artifact caused by the temperature dependence of background losses ignored earlier. Despite experimental uncertainties, the lack of strong temperature dependence in the low-temperature regime is consistent with the behavior of a nodeless superconductor, where thermal excitations of quasiparticles are exponentially suppressed. On the other hand, in the presence of gap nodes, $\sigma_2(T)-\sigma_2(0)$ is expected to scale with $T$ in pristine samples or $T^2$ with pairing-breaking scattering~\cite{hirschfeld_effect_1993}, neither of which is observed in our device~\cite{supplemental}.

To quantitatively evaluate the temperature dependence, we fit $\sigma_2$ to the following expression based on the Mattis-Bardeen theory in the dirty limit assuming a momentum-independent superconducting gap~\cite{mattis_theory_1958,dressel_electrodynamics_2002} :
\begin{equation}
\begin{split}
    \sigma_2(T) &= \frac{\eta \sigma_N}{\hbar \omega}\cdot  \int_{\text{max}(\Delta_T-\hbar\omega, -\Delta_T)}^{\Delta_T}\\
   &\frac{[1-2f_T(E+\hbar\omega)](E^2+\Delta_T^2+\hbar\omega E)}{\sqrt{(\Delta_T^2-E^2)[(E+\hbar\omega)^2-\Delta_T^2]}}dE.
\end{split}
\label{eq:MB2}
\end{equation}
This limit is justified because the optically measured normal-state scattering rate~\cite{mathew_roy_interlayer_2025} far exceeds the superconducting gap size. Here, $\sigma_N$ denotes the experimentally measured normal-state sheet conductance, $f_T$ is the Fermi function at $T$, and the superconducting gap is parameterized using the phenomenological interpolation formula~\cite{gross_anomalous_1986,prozorov_magnetic_2006}:
\begin{equation}
    \Delta_T = \Delta_0 \tanh\left({\alpha \sqrt{(T_\text{c} - T)/T}}\right).
\label{eq:gap}
\end{equation}
We also include a prefactor $\eta$ that describes the condensed fraction of spectral weight within the frequency window below the gap. The temperature evolution of $\sigma_1$ is excluded from this fit because of its large uncertainties below $T_\text{c}$.

The fitted curve, plotted in Fig.~\ref{Fig_Result}, agrees well with the experimental data. We find $2\Delta_0/k_\text{B}T_{\text{c}} = 3.9 (1)$ and $\alpha=2.4 (1)$. These values slightly exceed the Bardeen–Cooper–Schrieffer (BCS) predictions in the weak-coupling limit, but are comparable to those of moderately coupled superconductors such as Nb~\cite{supplemental}. The extracted $2\Delta_0/k_\text{B}T_{\text{c}}$ also matches the ratio of 4 estimated using the specific heat jump in bulk crystals~\cite{ribak_chiral_2020} and 3.8 from scanning tunneling spectroscopy (STS) measurements on the cleaved 1H surface~\cite{nayak_evidence_2021}. The fitting results reaffirm the nodeless nature of our device. Because the 4-layer sample necessarily contains a 1H-TaS$_2$ termination layer with broken mirror symmetry, our result does not support the scenario of surface nodal superconductivity proposed in Ref~\cite{nayak_evidence_2021}.

The extracted $\eta = 0.19 (1)$ is much smaller than unity, indicating a substantial fraction of uncondensed spectral weight. This echos bulk electronic specific heat ($C_\text{e})$ measurements~\cite{ribak_chiral_2020}, where the Sommerfeld coefficient $C_{\text{e}}/T$ remains finite in the low-temperature limit, signaling the existence of gapless states. However, the uncondensed spectral weight fraction in our measurement, $1-\eta$, is larger than the normal-metal fraction in $C_{\text{e}}/T$ from Ref.~\cite{ribak_chiral_2020}. This may originate from the difference between bulk crystals. In Ref~\cite{ribak_chiral_2020}, a small amount of Se is added in the crystal growth, while it is not included in the bulk crystals used here for exfoliation. The Se addition has been reported to increase the superconducting volume fraction~\cite{meng_extreme_2024}, but the underlying mechanism remains to be understood. Another factor to consider is that the gapless states may disproportionally affect these two measurements. For instance, a smaller effective mass associated with the gapless states would reduce their contributions to $C_{e}$ while enhancing their share of the $\sigma$ spectral weight.

We remark that while uncondensed spectral weight and residual gapless states are often observed in disordered nodal superconductors, they do not necessarily imply nodal superconductivity. In fact, theory suggests that in 4Hb-TaS$_2$, gapless Fermi pockets can coexist with fully gapped ones, when the pairing interaction is mostly intrapocket and the magnetic fluctuations in the 1T layer induce pair-breaking scattering~\cite{dentelski_robust_2021}. This scenario is supported by our observations in the 4-layer sample, as well as by specific heat~\cite{ribak_chiral_2020}, thermal conductivity~\cite{wang_evidence_2025}, and penetration depth~\cite{zhou_nodeless_2025} measurements on bulk crystals. In contrast, a nodal scenario fails to explain any of these experiments.

In summary, we have developed a cavity sensing method for measuring the microwave conductivity in atomically thin superconductors. Applied to a 4-layer 4Hb-TaS$_2$ sample, our measurement reveals nodeless superconductivity despite the presence of broken mirror symmetry. Our approach does not require ohmic or superconducting contacts, or any etching of the vdW heterostructure. As such, it can be readily applied to broad classes of 2D superconductors.

\medskip

\begin{acknowledgments}
\paragraph{Acknowledgments.} The microwave measurements were supported by the U.S. Department of Energy, Office of Science, Office of Advanced Scientific Computing Research Quantum Testbed Program under contract No. DE-AC02-05-CH11231. The 2D material device fabrication and DC transport measurements were supported by the U.S. Department of Energy, Office of Science, Office of Basic Energy Sciences, Materials Sciences and Engineering Division under contract No. DE-AC02-05-CH11231 within the van der Waals heterostructures program (KCWF16). The synthesis of bulk 4Hb-TaS$_2$ was supported by the U.S. Department of Energy, Office of Science, Basic Energy Sciences, Materials Sciences and Engineering Division under Contract No. DEAC02-05-CH11231 within the Quantum Materials program (KC2202). K.W. and T.T. acknowledge support from the JSPS KAKENHI (Grant Numbers 21H05233 and 23H02052), the CREST (JPMJCR24A5), JST and World Premier International Research Center Initiative (WPI), MEXT, Japan. T.C. acknowledges support from the National Science Foundation Graduate Research Fellowship Program (NSF GRFP) under Grant Nos. DGE 1752814 and DGE 2146752. H.-L.K. and R.Q. acknowledge support from the Kavli ENSI Graduate Student Fellowship. S.-D.C. acknowledges support from the Kavli ENSI Heising-Simons Junior Fellowship.
\medskip
\paragraph{Author contributions.} S.-D.C., T.C., and F.W. proposed and designed the experiment; T.C. performed the microwave measurements with supervision from D.I.S. and I.S.; H.-L.K., Q.W., and R.Q. performed the DC transport measurements; S.-D.C. developed the circuit model; S.-D.C., T.C., and Q.W. analyzed the data; H.-L.K., Q.W., and C.S. fabricated the 2D material devices; T.C., H.K., and Z.P. fabricated the superconducting resonators; L.P.C., R.P.D., and J.G.A. grew the 4Hb-TaS$_2$ crystals; T.T. and K.W. grew the hBN crystals; S.-D.C., T.C, and F.W. wrote the manuscript with input from all authors.
\medskip
\paragraph{Data availability.} The data underlying the figures in this work are openly available~\cite{chistolini_2026_dataset}.

\end{acknowledgments}

\appendix

\section{End Matter: Circuit Model}

\begin{figure*}
    \includegraphics[]{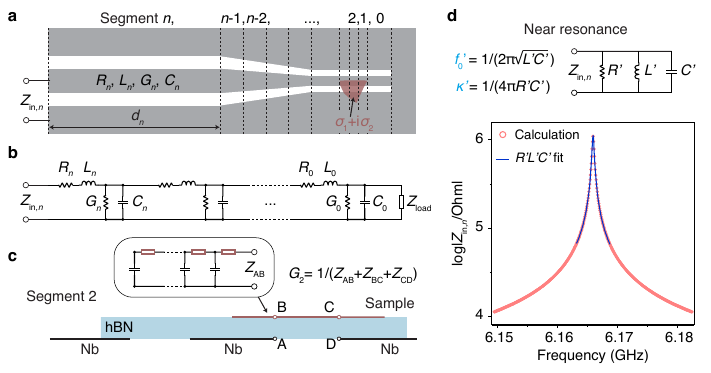}
    \caption{Circuit model mapping between $(f_0',\kappa')$ and $\sigma$. (a) Schematic of the sample resonator. The Nb and sample flake are shown in gray and red, respectively. The structure is discretized into $n+1$ segments, each with its own cross-sectional geometry. The meandered section following the taper (Fig.~\ref{Fig_1}(b)) is modeled as a straight segment (segment $n$), with an effective length $d_n$ that is not precisely known. Although the segments illustrated here are sparse (for clarity), the actual segmentation used is dense enough for our calculation to converge. (b) Circuit model of the structure in (a). $R_m$, $L_m$, $G_m$, and $C_m$ denote the resistance, inductance, admittance, and capacitance per unit length for segment~$m$, respectively. $Z_{\text{load}}$ is a load resistor that accounts for $\kappa_{\text{bkg,S}}$. $Z_{\text{in},n}$ represents the input impedance at the end of segment $n$. (c) Cross-sectional schematic using segment~2 as an example. $G_2$ is determined by the sum of impedance between points A and B, B and C, and C and D ($Z_{\text{AB}}$, $Z_{\text{BC}}$, and $Z_{\text{CD}}$). Inset: cross-sectional effective circuit for the overlap region between sample and Nb in segment~2, where $Z_{\text{AB}}$ equals the input impedance of the finite-length transmission line. (d) Bottom: $|Z_{\text{in},n}|$ calculated for a given set of $\sigma$, $d_n$, and $Z_\text{load}$, plotted on log scale as a function of frequency. Data are fit to the input impedance of an effective resonator circuit (top) with parameters $R'$, $L'$, and $C'$, which allows the extraction of $f_0'$ and $\kappa'$. The fitting range is adaptively set to approximately $40\kappa'$ centered at $f_0'$.}
    \label{Fig_Circuit}
\end{figure*}

Here we introduce the circuit model that maps between $f_0'$, $\kappa'$ and the sample conductance $\sigma = \sigma_1 + i\sigma_2$. We start by considering the sample resonator as shown in Fig.~\ref{Fig_Circuit}(a). To best model the tapered section and the irregular sample shape, we discretize the structure into many segments, each with their own cross-sectional geometry. $Z_{\text{in},m}$, the input impedance at the end of segment $m$, is related to $Z_{\text{in},m-1}$ by~\cite{pozar_microwave_2011}
\begin{equation}
    Z_{\text{in},m} = Z_{0,m} \cdot \frac{Z_{\text{in},m-1} + Z_{0,m} \text{tanh}(\gamma_m d_m)}{Z_{0,m} + Z_{\text{in},m-1} \text{tanh}(\gamma_m d_m)},
    \label{eq:Z_input_finite_line}
\end{equation}
where the characteristic impedance 
\begin{equation}
    Z_{0,m} = \sqrt{(R_m + i\omega L_m)/(G_m + i\omega C_m)},
\end{equation}
and propagation constant
\begin{equation}
\gamma_{m} = \sqrt{(R_m + i\omega L_m)(G_m + i\omega C_m)}.
\end{equation}
Here, $d_m$ is the length of segment~$m$, $\omega=2\pi f$ is the angular frequency, and $R_m$, $L_m$, $G_m$, and $C_m$ are the distributed resistance, inductance, admittance, and capacitance for the CPW segment~$m$, respectively. The phasor convention here follows Ref.~\cite{pozar_microwave_2011}.

Since the Nb is treated as perfect metal after background subtraction, we have $R_m = 0$. Moreover, as the widths of the CPW gap ($w_m$) and signal trace ($s_m$) are much larger than the Nb thickness and much smaller than the thickness of the Si substrate, the following expressions are used to calculate $C_m$ and $L_m$~\cite{simons_coplanar_2001}:
\begin{equation}
\begin{split}
    C_m &= 2\epsilon_0 (\epsilon_{\text{Si}}+1) K(k_m)/K(k_m'),\\
    L_m &= (\epsilon_{\text{Si}}+1)/(2c^2C_m).
\end{split}
\end{equation}
Here $k_m = s_m/(s_m + 2w_m)$, $k_m' = \sqrt{1-k_m^2}$, $K$ represents the complete elliptic integral of the first kind, $\epsilon_0$ is the vacuum permittivity,  $\epsilon_{\text{Si}} = 11.9$ is the dielectric constant of Si, and $c$ is the speed of light. 

The value of $G_m $ is zero in segments without the sample. For segments with the sample (such as segment 2 in Fig.~\ref{Fig_Circuit}(a)), we calculate $G_m$ using the cross-sectional circuit model developed in Ref.~\cite{gallagher_quantum-critical_2019} (see Supplemental Material~\cite{supplemental} for more details). As shown in the schematic in Fig.~\ref{Fig_Circuit}(c), the sample essentially connects the signal trace to the ground plane through "capacitive contacts" formed in regions overlapping with the Nb. The "contact" impedance $Z_{\text{AB}}$ between points A and B is the input impedance of the finite-length, open-ended transmission line shown in Fig.~\ref{Fig_Circuit}(c) inset:
\begin{equation}
    Z_{\text{AB}} = Z'_0 \coth\left(\gamma'l_m\right),
\end{equation}
where
\begin{equation}
    \begin{split}
        Z'_0 &= \sqrt{ t_{\text{hBN}} /\left(i \omega \epsilon_0 \epsilon_{\text{hBN}} \sigma \right)},\\
        \gamma'&=\sqrt{ i \omega \epsilon_0 \epsilon_{\text{hBN}}/\left( t_{\text{hBN} }\sigma\right) }.
    \end{split}
\end{equation}
Here $l_m$ represents the overlapping length between the flake and Nb on the left-hand side of the CPW gap in the cross-sectional view, $t_{\text{hBN}}\approx2.2~\text{nm}$ is the bottom hBN thickness, and $\epsilon_{\text{hBN}}\approx2.8$ is the hBN c-axis dielectric constant~\cite{chen_direct_2025}. Similarly, between points C and D, $Z_{\text{CD}} = Z'_0 \coth\left(\gamma'l'_m\right)$, where $l'_m$ denotes the overlapping length on the right-hand side. In addition, the impedance between points B and C is given by $Z_{\text{BC}}~=~w_m/\sigma$. These expressions allow $G_m~=~1/(Z_{\text{AB}}+Z_{\text{BC}}+Z_{\text{CD}})$ to be calculated in the presence of the sample.

With $R_m$, $L_m$, $G_m$, and $C_m$ determined for all $n+1$ segments, we can use Eq.~\ref{eq:Z_input_finite_line} iteratively to obtain the input impedance of the resonator, $Z_{\text{in}, n}$, as a function of $f$. In addition to $\sigma$, we introduce two more free parameters in this calculation. The first is a load resistance $Z_{\text{load}}$ before segment 0 (Fig.~\ref{Fig_Circuit}(b)), which we use to account for the background loss in the system. The second is the length of the last segment $d_n$ (Fig.~\ref{Fig_Circuit}(a)), which is not precisely known because of the meandering design and the coupling structure to the feedline in addition to fabrication uncertainties. For a given set of $\sigma_1$,  $\sigma_2$, $Z_{\text{load}}$, and $d_n$ in the physically relevant range, the calculated $|Z_{\text{in},n}(f)|$ near its resonance can be well fit (see Fig.~\ref{Fig_Circuit}(d) for example) using the expression of a parallel-RLC resonator~\cite{pozar_microwave_2011}:
\begin{equation}
    |Z_{\text{in},n}| = \frac{1}{\left|1/R' + 1/(2\pi ifL')+ 2\pi ifC'\right|}.
\end{equation}
$f_0'$ and $\kappa'$ can thus be obtained from the fitting parameters $R'$, $L'$, and $C'$ as the following:
\begin{equation}
    \begin{aligned}
        f_0' &= \frac{1}{2\pi \sqrt{L'C'}},\\
        \kappa' &= \frac{1}{4\pi R'C'}.
    \end{aligned}
\end{equation}
This completes the numerical mapping between $(f_0', \kappa')$ and $(\sigma_1, \sigma_2)$ at given $Z_{\text{load}}$ and $d_n$. We have verified that in the large-$\sigma$ regime where cavity perturbation theory applies, our circuit model produces results consistent with the theory.

To extract $\sigma$, we notice that in the normal state, $\sigma_2 \approx 0$ at few-GHz frequencies. If either $Z_{\text{load}}$ or $\sigma_1$ is known, the other can be determined along with $d_n$ from the measured $(f_0', \kappa')$ values. Since $d_n$ and $Z_{\text{load}}$ are approximately temperature independent, they can then be used to convert $(f_0', \kappa')$ at every temperature to $(\sigma_1, \sigma_2)$. In practice, the exact value of $Z_{\text{load}}$ or normal-state $\sigma_1$ is not available in our device. Nevertheless, $Z_{\text{load}}$ has a known upper bound, because $\kappa_{\text{bkg,S}}$, which increases with $Z_{\text{load}}$, cannot exceed the measured $\kappa'$ in the low-temperature limit (Fig.~\ref{Fig_2}(d)). Moreover, $Z_{\text{load}}$ must be non-negative, so its range is bounded from both sides. These bounds in turn allow us to determine the corresponding ranges for $\sigma_1$ and $\sigma_2$ at each temperature. In future studies, this uncertainty may also be eliminated by directly measuring the normal-state $\sigma_1$ via DC transport.

\bibliography{reference}

\end{document}


\newcommand{\lbnlafil}{Applied Mathematics and Computational Research Division, Lawrence Berkeley National Laboratory, Berkeley, California 94720, USA}
\newcommand{\foundryafil}{Molecular Foundry Division, Lawrence Berkeley National Laboratory, Berkeley, California 94720, USA}
\newcommand{\lbnlmatafil}{Materials Science Division, Lawrence Berkeley National Laboratory, Berkeley, California 94720, USA}
\newcommand{\qnlafil}{Quantum Nanoelectronics Laboratory, Department of Physics, University of California, Berkeley, California 94720, USA}
\newcommand{\berkeleyphysafil}{Department of Physics, University of California, Berkeley, California 94720, USA}
\newcommand{\KavliENSI}{Kavli Energy NanoScience Institute at the University of California, Berkeley and the Lawrence Berkeley National Laboratory, Berkeley, California 94720, USA}
\newcommand{\ufl}{Department of Physics, University of Florida, Gainesville, Florida 32611, USA}

\newcommand{\nimsa}{Research Center for Materials Nanoarchitectonics, National Institute for Materials Science, 1-1 Namiki, Tsukuba 305-0044, Japan}

\newcommand{\nimsb}{Research Center for Electronic and Optical Materials, National Institute for Materials Science; 1-1 Namiki, Tsukuba 305-0044, Japan}

\title{Supplemental Material for\\Contactless cavity sensing of superfluid stiffness in atomically thin 4Hb-TaS$_2$}
\author{Trevor Chistolini}
\thanks{These authors contributed equally}
\affiliation{\berkeleyphysafil}
\affiliation{\lbnlafil}

\author{Ha-Leem Kim}
\thanks{These authors contributed equally}
\affiliation{\berkeleyphysafil}
\affiliation{\lbnlmatafil}
\affiliation{\KavliENSI}

\author{Qiyu Wang}
\thanks{These authors contributed equally}
\affiliation{\berkeleyphysafil}

\author{Su-Di Chen}
\thanks{These authors contributed equally}
\thanks{Corresponding author}
\email{sudichen@ufl.edu}
\affiliation{\berkeleyphysafil}
\affiliation{\lbnlmatafil}
\affiliation{\KavliENSI}
\affiliation{\ufl}

\author{Luke Pritchard Cairns}
\affiliation{\berkeleyphysafil}

\author{Ryan Patrick Day}
\affiliation{\berkeleyphysafil}

\author{Collin Sanborn}
\affiliation{\berkeleyphysafil}

\author{Hyunseong Kim}
\affiliation{\berkeleyphysafil}

\author{Zahra Pedramrazi}
\affiliation{\lbnlafil}

\author{Ruishi Qi}
\affiliation{\berkeleyphysafil}
\affiliation{\lbnlmatafil}
\affiliation{\KavliENSI}

\author{Takashi Taniguchi}
\affiliation{\nimsa}

\author{Kenji Watanabe}
\affiliation{\nimsb}

\author{James G. Analytis}
\affiliation{\berkeleyphysafil}
\affiliation{\lbnlmatafil}

\author{David I. Santiago}
\affiliation{\lbnlafil}

\author{Irfan Siddiqi}
\affiliation{\berkeleyphysafil}
\affiliation{\lbnlafil}
\affiliation{\KavliENSI}

\author{Feng Wang}
\thanks{Corresponding author}
\email{fengwang76@berkeley.edu}
\affiliation{\berkeleyphysafil}
\affiliation{\lbnlmatafil}
\affiliation{\KavliENSI}

\date{\today}

\maketitle

\section{Device fabrication}
The Nb resonators are fabricated using a wafer-scale process. We begin with a high-resitivity, 6-inch Si wafer ($\geq10,000~\Omega\cdot \text{cm}$), which is $675~\mu\text{m}$ thick. The wafer is initially cleaned using a Piranha etch and a buffered oxide etch (BOE), and then a $200$-nm thick Nb film is sputtered. For lithography, we use $1.0~\mu\text{m}$ of AZ MiR 701 photoresist, and then optically write the resonators, transmission line, and other ground plane features. After development, the Nb is etched using a chlorine-based process. The wafer is diced into $5~\text{mm}\times5~\text{mm}$ chips.

For the feedline, the inner conductor width is $30~\mu\text{m}$ and the gap between the inner and outer conductors is $17.5~\mu\text{m}$. For the resonators, the width and gap are $20~\mu\text{m}$ and $10~\mu\text{m}$, respectively; these are gradually tapered down to $8~\mu\text{m}$ and $5~\mu\text{m}$ near the sample position. The coupling segment to the feedline is $150~\mu\text{m}$ long, with the inner conductors of the resonator and the feedline separated by $27.5~\mu\text{m}$. This design achieves relatively strong coupling with a measured external quality factor ($Q_e$) around $3.5\times10^3$, allowing $f_0/2Q_{i}$ over a large range to be reliably extracted.

Single crystals of 4Hb-TaS$_2$ are grown by a two-step procedure---a precursor step and then chemical vapor transport (CVT). For the precursor, Ta (ThermoScientific, $99.98\%$ purity) and S (ThermoScientific, $99.999\%$ purity) are combined in a molar ratio of 1:2.2 (Ta:S), then loaded in an alumina crucible and sealed in a quartz tube with 200 Torr Ar gas. The tube is heated to 950 $\degree\text{C}$, held for 12 hours and then quenched in ice water. For the second step, the precursor is ground with a mortar and pestle, then loaded with 3 mg/cm$^3$ iodine in a 21 cm long quartz tube, evacuated, and placed in a horizontal two-zone furnace. The two zones are held at 780 $\degree\text{C}$ (source) and 680 $\degree\text{C}$ (growth) for 27 days, then allowed to cool to room temperature naturally. Single crystals are then collected from the growth end.

The single crystals are subsequently exfoliated onto SiO$_2$/Si chips inside a nitrogen glovebox. Crystals without Se are used because they are found to be easier to exfoliate. Under a microscope, suitable flakes are identified based on their optical contrast. hBN flakes are also exfoliated. Using a bisphenol-A polycarbonate (PC) stamp, we sequentially pick up the top hBN, the few-layer 4Hb-TaS$_2$, and the bottom hBN. The assembled stack is then released onto the Nb resonator and the PC residue is removed using chloroform.

\begin{figure}
    \centering
    \includegraphics[]{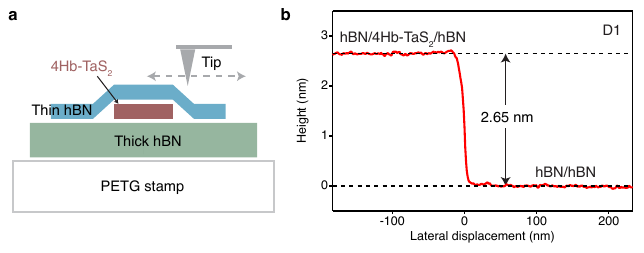}
    \caption{Determination of sample layer number. (a) Schematic of sample geometry in AFM measurements. The stack is picked up from the resonator using a PETG stamp such that the thin hBN is exposed on the surface. The tip is scanned on the hBN surfaces. (b) Representative contact-mode line scan across the 4Hb-TaS$_2$ edge, measured on D1, the device studied in the main text. The step height $h = t_{\text{S}}+2d_{\text{S-hBN}}-d_{\text{hBN-hBN}}$, where $t_{\text{S}}$ is the sample thickness, $d_{\text{S-hBN}}$ is the van der Waals (vdW) gap between the sample and hBN, and $d_{\text{hBN-hBN}}$ is the vdW gap between two hBN flakes. Using measurements from an hBN/monolayer MoSe$_2$/hBN stack, we estimate $2d_{\text{S-hBN}}-d_{\text{hBN-hBN}} \approx 0.3\text{ nm}$. The measured step height $h\approx 2.65\text{ nm}$ therefore yields sample thickness $t_{\text{S}}\approx 2.4\text{ nm}$, consistent with the expected thickness of 2.36~nm~\cite{di_salvo_preparation_1973} for 4-layer 4Hb-TaS$_2$.}
    \label{FigS_AFM}
\end{figure}

After the microwave measurements, the stack is picked up from the resonator using a glycol-modified polyethylene terephthalate (PETG) stamp such that the thin bottom hBN is now exposed (Fig.~\ref{FigS_AFM}(a)). The thickness of the 4Hb-TaS$_2$ flake as well as that of the thin hBN is then measured using an atomic force microscope (AFM, Asylum Research Cypher S), and the layer number is determined accordingly (see Fig.~\ref{FigS_AFM}(b) for example).

The DC transport devices are fabricated using similar procedures. For device D7, few-layer graphite flakes are used as contacts and the 4Hb-TaS$_2$ is encapsulated on both sides by hBN. For D8 and D9, pre-patterned Au contacts are used and the 4Hb-TaS$_2$ is covered by hBN on top and in contact with the SiO$_2$/Si substrate on the bottom side.

\section{Microwave measurements}
We measure all of the resonator devices in an adiabatic demagnetization refrigerator (ADR), using a standard microwave setup. A vector network analyzer (VNA) is used to measure the scattering parameter $S_{21}$ as a function of frequency near the fundamental modes of the resonators. Further details on the setup are presented in Ref.~\cite{chistolini_performance_2024}. The dataset for each device is collected during a single cool down, with both the sample and the reference resonators being measured together at each temperature.

\begin{figure}
    \centering
    \includegraphics[]{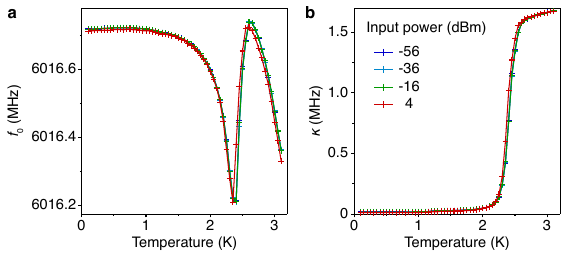}
    \caption{Effect of input microwave power on measurements. (a) $f_0$ of the sample resonator in D1 measured at varying input power levels. (b) Same as (a) for $\kappa$. The corresponding power levels of the VNA output at room temperature are labeled in (b). Curves below -16 dBm all overlap with each other, indicating negligible nonlinear effects. The deviation only becomes notable at 4 dBm, likely because of heating and quasiparticle generation effects.}
    \label{FigS_power}
\end{figure}

At each temperature, the resonances are measured across a range of power levels. In Fig.~\ref{FigS_power}, we show the temperature evolutions of $f_0$ and $\kappa$ at selected power levels. In the low-power regime, we observe a converging behavior where $f_0$ and $\kappa$ do not change with power. This indicates negligible effects from heating or quasiparticle generation. We focus on this linear regime to study the equilibrium properties of the sample in the main text, where the data are taken with a VNA input power of -56~dBm.

\section{Extracting resonator properties from $S_{21}$ spectra}\label{SI_scraps}

To extract $f_0$ and $\kappa \coloneq f_0/2Q_i$ for each resonator, we fit the complex $S_{21}$ spectra using the Superconducting Analysis and
Plotting Software (scraps) package~\cite{carter_scraps_2017}. The model is given by:
\begin{equation}
\begin{gathered}
    S_{21}(f) = g(f)e^{i\phi(f)}\cdot \frac{Q_e + i Q_e Q_i (2x + 2\delta f/\tilde{f_0})}{Q_i+Q_e + 2iQ_iQ_ex},\\
    g(f) = g_0 + g_1x+g_2 x^2,\\
    \phi(f) = \phi_0+\phi_1 x,\\
    \tilde{f_0} = f_0 + \delta f,\\
    x = (f-\tilde{f_0})/\tilde{f_0}.
\label{eq_scraps}
\end{gathered}
\end{equation}
Here, $g(f)e^{i\phi(f)}$ is the environment term that accounts for the attenuation, amplification, and phase shift in the measurement system, the electronic delay and damping in the cables, as well as slow-varying transmission-line resonances~\cite{probst_efficient_2015, carter_scraps_2017}. The last term in $S_{21}(f)$ models an asymmetric resonance which takes impedance mismatch into account~\cite{khalil_analysis_2012, geerlings2013improving, carter_scraps_2017}.

\begin{figure}
    \centering
    \includegraphics[]{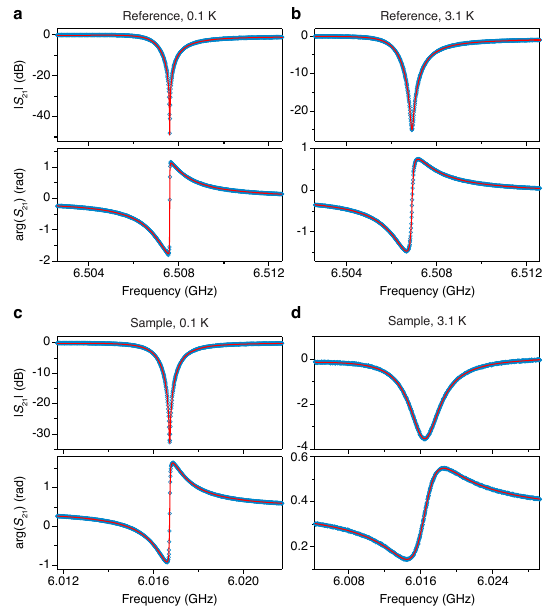}
    \caption{Representative fit curves for extracting $f_0$ and $\kappa$. Data (blue circles) are measured in D1. The corresponding resonator and temperature information is labeled above each pair of panels. The complex $S_{21}$ as a function of frequency is fitted using the scraps package~\cite{carter_scraps_2017} (see section~\ref{SI_scraps} for details) and the results are plotted in red.}
    \label{FigS_FittingDetails}
\end{figure}

In Fig.~\ref{FigS_FittingDetails}, we plot the $S_{21}$ data and the corresponding fit curves from device 1 (D1, studied in the main text) for both the reference and the sample resonators at the lowest and highest measured temperatures. The agreement between the data and fit curves is excellent. In the sample resonator, the $1\sigma$ error propagated from the fit is around 0.1~kHz for both $f_0$ and $\kappa$ at base temperature and increases to around 1.5~kHz at 3.1~K. In the reference resonator, the $1\sigma$ error remains around 0.1~kHz for all temperatures. These fitting errors are negligible in comparison to the temperature-induced changes in $f_0$ and $\kappa$ (on the scale of hundreds of kHz) and are thus ignored in our analysis.

\section{Sample-to-sample variation}
\begin{figure}
    \centering
    \includegraphics[]{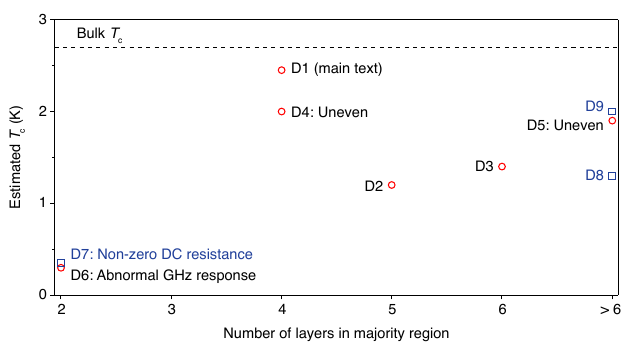}
    \caption{Distribution of $T_{\text{c}}$ among all devices studied plotted versus the number of layers. Dashed line marks bulk $T_{\text{c}}$. Red circles and blue squares represent microwave and DC transport devices, respectively. D1 is the device studied in the main text. D4 and D5 have nonuniform, minority regions with different thickness and are thus labeled as uneven. D6 and D7 show transition-like behavior but remain resistive in the low-temperature limit. See Figs.~\ref{FigS_AD} and \ref{FigS_DC} for more details.}
    \label{FigS_SS}
\end{figure}

\begin{figure}
    \centering
    \includegraphics[width = \textwidth]{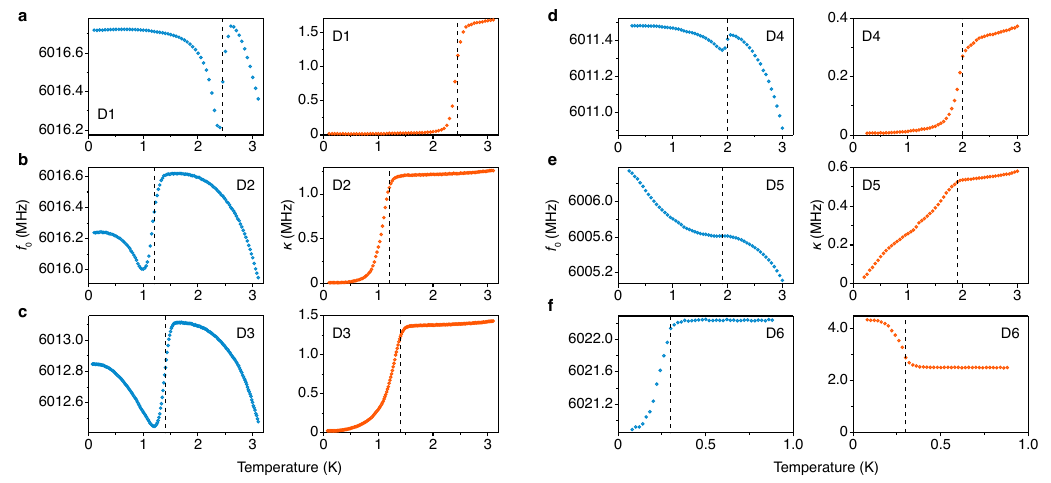}
    \caption{$f_0$ and $\kappa$ of the sample resonators from all microwave devices. The corresponding device name is labeled in each panel. Vertical dashed lines mark the estimated $T_{\text{c}}$.}
    \label{FigS_AD}
\end{figure}

\begin{figure}
    \centering
    \includegraphics[]{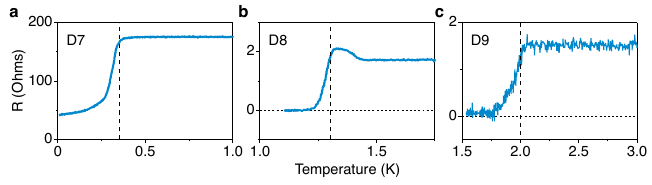}
    \caption{Resistance versus temperature from all DC transport devices. Data are obtained from 4-contact measurements using standard lock-in techniques. Vertical dashed lines mark the estimated transition temperatures. The corresponding device name is labeled in each panel.}
    \label{FigS_DC}
\end{figure}

We measure 6 microwave devices and 3 transport devices (see Fig.~\ref{FigS_SS} for summary and Figs.~\ref{FigS_AD} and \ref{FigS_DC} for details). Except for D1 studied in the main text, significant and irregular suppression of $T_{\text{c}}$ from the bulk value is observed in all other devices. This variation could arise from inhomogeneities in the bulk crystals or the degradation of few-layer samples in the imperfect and time-varying glovebox environment. The same effect may also create a heterogeneous distribution of local $T_{\text{c}}$ within each flake, smearing the temperature evolution of $f_0$ and $\kappa$.

In D1, D2, and D4 (Fig.~\ref{FigS_AD}(a), (b), and (d)), we observe qualitatively similar behaviors despite the $T_\text{c}$ variation. Importantly, $f_0 (T)$ flattens in the low-temperature regime, indicating the presence of a nodeless superconducting gap (see Section~\ref{SICP} for more discussions).

Based on the microwave data alone, we cannot rule out the possible existence of nodes in D3 (Fig.~\ref{FigS_AD}(c)), where the data below 0.5 K could be consistent with $\sigma_2(0)-\sigma_2(T) \propto T^2$, and D5 (Fig.~\ref{FigS_AD}(e)), where $f_0$ evolves linearly with $T$ down to the base temperature. However, these behaviors could also arise from sample inhomogeneity instead, because averaging over a distribution of local $T_{\text{c}}$ may create an apparent temperature dependence in $\sigma_2(T)$ and thus $f_0(T)$ at low temperature, even if the sample is nodeless everywhere. Importantly, the flakes with lower $T_c$ lie further within the dirty limit, where the scattering rate~\cite{mathew_roy_interlayer_2025} far exceeds the superconducting gap size. Nodal superconductivity is highly unlikely to survive in this limit.

In addition, we note that D6 shows increased dissipation at lower temperature (Fig.~\ref{FigS_AD}(f)), opposite to the other devices. This is likely associated with a failed superconducting state, where the flake remains dissipative instead of inductive below the transition temperature (Fig.~\ref{FigS_DC}(a)).

\section{Mattis-Bardeen fit for reference resonator}
In the main text, the temperature evolution of the reference resonator is modeled as
\begin{equation}
\begin{split}
f_{0,\text{R}}(T) = f_{00,\text{R}} + g_{\text{R}}\cdot\text{Im}Z_{\text{Nb}}(T),\\
\kappa_{\text{R}}(T) = \kappa_{\text{bkg,R}} + g_{\text{R}}\cdot\text{Re}Z_{\text{Nb}}(T).
\label{eq_ref_SI}
\end{split}
\end{equation}
We express the surface impedance of Nb using its complex conductivity $\sigma = \sigma_1+i\sigma_2$:
\begin{equation}
    Z_{\text{Nb}} = \left(\frac{\mu_0\omega}{2}\frac{\sqrt{\sigma_1^2+\sigma_2^2}-\sigma_2}{\sigma_1^2+\sigma_2^2}\right)^{1/2} - i \left(\frac{\mu_0\omega}{2}\frac{\sqrt{\sigma_1^2+\sigma_2^2}+\sigma_2}{\sigma_1^2+\sigma_2^2}\right)^{1/2}.
\label{eq:Nb_SI}
\end{equation}
According to the Mattis-Bardeen theory, we have:
\begin{equation}
\begin{split}
    \sigma_1(\omega,T) = \frac{2\sigma_N}{\hbar \omega}\cdot
    \int_{\Delta_{T}}^{\infty}\frac{[f_T(E)-f_T(E+\hbar\omega)](E^2+\Delta_T^2+\hbar\omega E)}{\sqrt{(E^2-\Delta_T^2)[(E+\hbar\omega)^2-\Delta_T^2]}}dE\\
    +\frac{\sigma_N}{\hbar \omega}\cdot  \int_{\Delta_T-\hbar\omega}^{-\Delta_T}
   \frac{[1-2f_T(E+\hbar\omega)](E^2+\Delta_T^2+\hbar\omega E)}{\sqrt{(E^2-\Delta_T^2)[(E+\hbar\omega)^2-\Delta_T^2]}}dE,
\end{split}
\label{eq:MB1_SI}
\end{equation}
\begin{equation}
  \sigma_2(T) = \frac{\sigma_N}{\hbar \omega}\cdot  \int_{\text{max}(\Delta_T-\hbar\omega, -\Delta_T)}^{\Delta_T}
   \frac{[1-2f_T(E+\hbar\omega)](E^2+\Delta_T^2+\hbar\omega E)}{\sqrt{(\Delta_T^2-E^2)[(E+\hbar\omega)^2-\Delta_T^2]}}dE.
\label{eq:MB2_SI}
\end{equation}
The second term in Eq.~\ref{eq:MB1_SI} should be set to 0 for $\hbar\omega < 2\Delta_T$, which is always the case in our measurement range. The superconducting gap of Nb is assumed to follow the BCS-like phenomenological interpolation formula~\cite{gross_anomalous_1986,prozorov_magnetic_2006}:
\begin{equation}
    \Delta_T = \Delta_0 \tanh\left({\alpha \sqrt{(T_\text{c} - T)/T}}\right).
\label{eq:gap_SI}
\end{equation}
In the weak-coupling BCS limit, $2\Delta_0/k_{\text{B}}T_{\text{c}}\approx 3.53 $ and $\alpha \approx 1.74$. Both values are expected to increase with stronger electron-boson coupling.

Eqs.~\ref{eq_ref_SI}--\ref{eq:gap_SI} allow us to simultaneously fit the experimentally measured $f_{0,\text{R}}$ and $\kappa_\text{R}$ as functions of $T$. The independent fitting parameters are $f_{00,\text{R}}$, $\kappa_{\text{bkg,R}}$, $g_{\text{R}}/\sqrt{\sigma_{\text{N}}}$, $\alpha$, $T_\text{c}$, and $2\Delta_0/k_{\text{B}}T_{\text{c}}$. We obtain a $T_\text{c}$ of 9~K and a $2\Delta_0/k_\text{B} T_\text{c}$ around 3.9, in good agreement with literature values for Nb~\cite{townsend_investigation_1962}. $\alpha$ is found to be near 2.5, consistent with the fact that Nb is a moderately coupled superconductor.

\section{Frequency dependence of niobium response}

\begin{figure}
    \centering
    \includegraphics[]{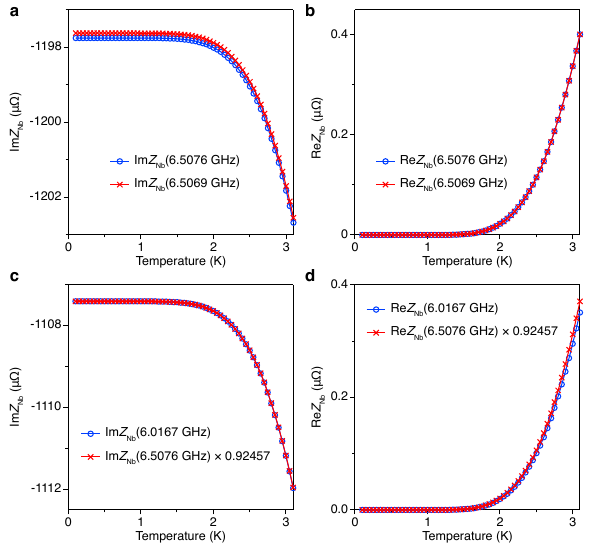}
    \caption{Calculated temperature evolution of $Z_{\text{Nb}}$ at frequencies relevant to our measurements. The calculation uses Eqs.~\ref{eq:Nb_SI}-\ref{eq:gap_SI} with parameters $T_\text{c} = 9\text{~K}$, $2\Delta_0/k_\text{B} T_\text{c}=3.9$, and $\alpha=2.5$ obtained from fitting. A realistic $\sigma_{\text{N}} = 2\times10^8\text{~S/m}$ is used. (a) $\text{Im}Z_{\text{Nb}}$ at 6.5076~GHz (blue) and 6.5069~GHz (red), corresponding to the resonant frequencies of the reference resonator at 0.1~K and 3.1~K, respectively. (b) $\text{Re}Z_{\text{Nb}}$ for the same frequencies. The $Z_{\text{Nb}}$ difference is negligible between these two frequencies. (c) $\text{Im}Z_{\text{Nb}}$ at 6.0167~GHz (blue), the resonant frequency of the sample resonator. The 6.5076~GHz curve from (a), rescaled by a factor of 0.92457 (the frequency ratio), is plotted in red for comparison. (d) Same as (c) for $\text{Re}Z_{\text{Nb}}$. After rescaling, the curves show negligible difference, justifying the omission of the frequency dependence in $Z_{\text{Nb}}$ in our analysis. The rescaling constant is implicitly absorbed into the ratio $g_{\text{S}}/g_\text{R}$ in the main text.}
    \label{FigS_Nb_fdep}
\end{figure}

We do not explicitly consider the frequency dependence of $Z_{\text{Nb}}$ in the main text. This section provides the justification for our approximation.

In our measurements, Nb is deep in the superconducting state. In this regime, $\sigma_1\ll\sigma_2$ and $\sigma_2\propto n_\text{s}(T)/f$ at microwave frequencies, where $n_\text{s}$ represents the superfluid density in Nb. Consequently, $Z_{\text{Nb}}$ is dominated by its imaginary part, which scales as $\text{Im}Z_{\text{Nb}}\propto f/\sqrt{n_{\text{s}}}$. The real part is suppressed by a factor of $-\sigma_1/(2\sigma_2)$ with respect to the imaginary part and scales as $\text{Re}Z_{\text{Nb}}\propto f^2 n_{\text{s}}^{-\frac{3}{2}}\sigma_1$.

For each resonator, the relative frequency shift in the measured temperature range is within $\sim10^{-4}$. Compared to the changes arising from the temperature dependence of $n_{\text{s}}$ and $\sigma_1$, this frequency shift induces a negligible change in $Z_{\text{Nb}}$, as shown in Fig.~\ref{FigS_Nb_fdep}(a) and (b). Therefore, the frequency dependence of $Z_{\text{Nb}}$ can be safely ignored in the analysis of a single resonator.

When analyzing the $f_0$ and $\kappa$ data of the sample resonator, we subtract out the background Nb contribution $(g_{\text{S}}/g_\text{R})\cdot Z_{\text{Nb}}$. Here $Z_{\text{Nb}}$ is obtained at the resonant frequency of the reference resonator, which differs from that of the sample resonator by approximately $8\%$. This procedure is justified because $Z_{\text{Nb}}$ is dominated by $\text{Im}Z_{\text{Nb}}\propto f/\sqrt{n_{\text{s}}}$, and the frequency scaling is implicitly accounted for by the ratio $g_{\text{S}}/g_\text{R}$ determined in Fig.~2(b). Although $\text{Re}Z_{\text{Nb}} \propto f^2$, its contribution to $Z_{\text{Nb}}$ is small enough, and the frequencies of the sample and the reference resonators are close enough, such that the error introduced is negligible, as demonstrated in Fig.~\ref{FigS_Nb_fdep}(d).

\section{Qualitative understanding of temperature evolution}\label{SICP}
\begin{figure}
    \includegraphics[]{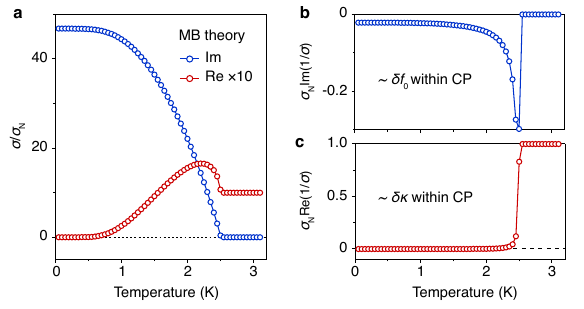}
    \caption{Typical temperature evolution of $1/\sigma$ in a superconductor. (a) Simulated microwave conductivity $\sigma$ in a typical BCS superconductor. Data are normalized by the normal state conductivity $\sigma_{\text{N}}$. The simulation uses Eqs.~\ref{eq:MB1_SI}-\ref{eq:gap_SI} with the following parameters: $T_{\text{c}}=2.5~\text{K}$, $2\Delta_0/k_{\text{B}}T_{\text{c}}=3.5$, $\alpha=1.74$, and $\omega/2\pi = 6.2~\text{GHz}$. (b) and (c), imaginary and real parts of $\sigma_\text{N}/\sigma$ as functions of temperature, which qualitatively reproduce the measured temperature evolutions of $f_0'$ and $\kappa'$, respectively.}
    \label{FigS_CP}
\end{figure}
Here, we provide a qualitative picture for understanding the temperature evolution of $f_0'$ and $\kappa'$ from a cavity perturbation (CP) perspective. If the 4Hb-TaS$_2$ flake is conducting enough, expressions similar to Eqs.~\ref{eq_ref_SI} can also be used to describe the impact of the flake on the sample resonator. Because the flake is atomically thin, the surface impedance should be replaced by $1/\sigma$, where $\sigma$ denotes the flake's sheet conductance. As shown in Fig.~\ref{FigS_CP}, the temperature evolution of $1/\sigma$ in a BCS superconductor can qualitatively reproduce the characteristic lineshapes of $f_0'(T)$ and $\kappa'(T)$ without fine tuning.

Importantly, in the low-temperature regime, since $\delta f_0\propto \text{Im}(1/\sigma)\approx -1/\sigma_2$, a flat $\delta f_0 (T)$ directly indicates the lack of temperature dependence in $\sigma_2 (T)$ and thus the absence of gap nodes. However, at temperatures near or above $T_\text{c}$, the actual $\sigma$ of 4Hb-TaS$_2$ may not be large enough for a strictly linear relationship between $1/\sigma$ and $\kappa'+if_0'$ to apply. Thus, a circuit model is still needed to quantitatively extract $\sigma$ from the measured data.

\section{Cross-sectional circuit model}
\begin{figure}
    \centering
    \includegraphics[]{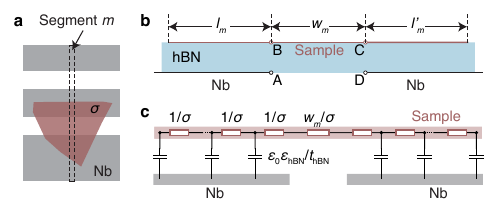}
    \caption{Cross-sectional circuit model. (a) Top view of the CPW near the 2D flake (red). (b) Cross-sectional schematic using segment~$m$ in (a) as an example. $l_m$ and $l'_m$ mark the overlapping length between the sample and Nb on the left and right, respectively. The ground plane farther away from the sample is not shown. (c) The effective circuit for (b). The overlapping regions are modeled as transmission lines, where the impedance per unit length is $1/\sigma$ and capacitance per unit length is $\epsilon_0\epsilon_{\text{hBN}}/t_{\text{hBN}}$.}
    \label{FigS_cross_sectional}
\end{figure}

We use the cross-sectional circuit model developed in Ref.~\cite{gallagher_quantum-critical_2019} to calculate the distributed admittance in the CPW segments containing the sample. Here we explain the model in more detail using segment~$m$ in Fig.~\ref{FigS_cross_sectional}(a) as an example.

The flake, which spans across the CPW gap, introduces a distributed admittance between the inner conductor and ground plane of the CPW:
\begin{equation}
G_m = 1/(Z_{\text{AB}}+Z_{\text{BC}}+Z_{\text{CD}}). 
\label{eq_cl0}
\end{equation}
As illustrated in Fig.~\ref{FigS_cross_sectional}(b) and (c), the impedance between points A and B ($Z_{\text{AB}}$) is the input impedance of the cross-sectional transmission line formed by the sample and Nb in the left overlapping region. This transmission line has length $l_m$ and is open-terminated, therefore~\cite{pozar_microwave_2011}
\begin{equation}
    Z_{\text{AB}} = Z'_0 \coth\left(\gamma'l_m\right).
    \label{eq_cl1}
\end{equation}
Here,  $Z'_0$ and $\gamma'$ are the characteristic impedance and propagation constant of the cross-sectional transmission line, determined by the distributed impedance ($R' + i\omega L'$) and  distributed capacitance ($C'$):
\begin{equation}
    Z'_{0} = \sqrt{(R' + i\omega L')/(i\omega C')},
    \label{eq_cl2}
\end{equation}
\begin{equation}
    \gamma' = \sqrt{(R' + i\omega L')(i\omega C')}.
    \label{eq_cl3}
\end{equation}
The distributed impedance is dominated by the sample's sheet resistance and kinetic inductance, as the geometrical inductance and Nb's kinetic inductance are negligible in comparison. Thus,
\begin{equation}
    R' + i\omega L'\approx 1/\sigma.
    \label{eq_cl4}
\end{equation}
The distributed capacitance is set by the hBN between the sample and Nb:
\begin{equation}
    C' = \epsilon_0 \epsilon_{\text{hBN}}/t_{\text{hBN}}.
    \label{eq_cl5}
\end{equation}
Similar to Eq.~\ref{eq_cl1}, the impedance between points C and D is
\begin{equation}
    Z_{\text{CD}} = Z'_0 \coth\left(\gamma'l'_m\right).
    \label{eq_cl6}
\end{equation}
In addition, the impedance between points B and C from the sample inside the gap is
\begin{equation}
    Z_{\text{BC}} = w_m/\sigma.
    \label{eq_cl7}
\end{equation}
Eqs.~\ref{eq_cl0}--\ref{eq_cl7} complete the expression for $G_m$. In segments without the sample, $G_m = 0$.

\section{Power-law fitting of low-temperature $\sigma_2$}

\begin{figure}
    \centering
    \includegraphics[]{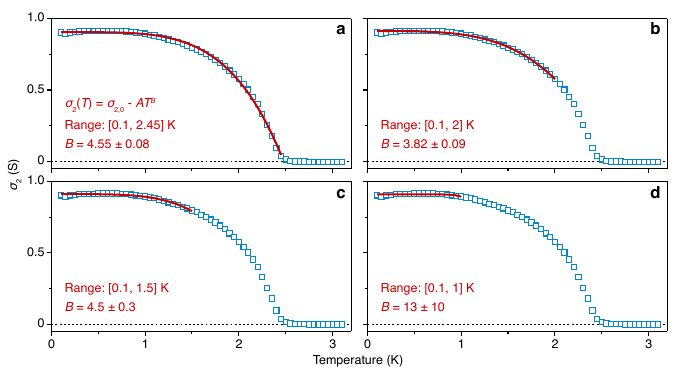}
    \caption{Temperature evolution of $\sigma_2$ (blue) fitted to the power law $\sigma_2(T) = \sigma_{2,0} - AT^B$, where $\sigma_{2,0}$, $A$, and $B$ are fitting parameters. The results are plotted in red. The fitting range and extracted exponent $B$ are labeled in each panel. The $B$ values are inconsistent with the existence of gap nodes.}
    \label{FigS_PowerExponent}
\end{figure}

In Fig.~\ref{FigS_PowerExponent} we fit the low-temperature $\sigma_2$ data from D1 to the power law $\sigma_2(T) = \sigma_{2,0} - AT^B$. For 2D nodal superconductors, the exponent $B$ is expected to be 1 in the clean limit or 2 in the presence of pair-breaking scattering. Although the fitting results depend on the temperature range chosen, the extracted values for $B$ are generally around 4 or larger, indicating nodeless superconductivity.

\section{Possible effect from niobium oxide layer}
\begin{figure}
    \includegraphics[]{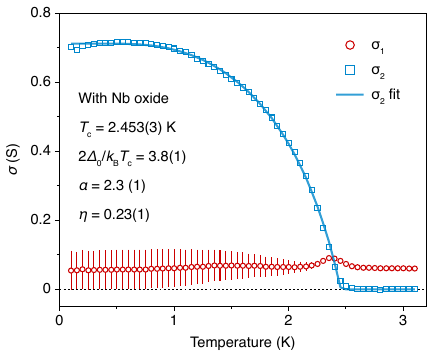}
    \caption{Temperature evolution of sheet conductance $\sigma$ at 6 GHz for the 4-layer 4Hb-TaS$_2$ flake, obtained using the circuit model with the Nb oxide layer taken into account. While the absolute conductivity values are approximately $20\%$ lower than those in Fig.~3, the temperature-dependent lineshape is nearly identical and the extracted fit parameters are consistent with those in the main text.}
    \label{Fig_Result_NbOxide}
\end{figure}

A native oxide layer can exist on the surface of Nb, with typical thickness $t_{\text{oxide}}\sim 5\text{~nm}$~\cite{altoe_localization_2022} and dielectric constant $\epsilon_{\text{oxide}}\sim30$~\cite{fuschillo_dielectric_1975}. Taking this additional dielectric layer into account, Eq.~\ref{eq_cl5} becomes 
\begin{equation}
    C' = \frac{\epsilon_0}{  {t_{\text{hBN}}}/{\epsilon_{\text{hBN}}} + {t_{\text{oxide}}}/{\epsilon_{\text{oxide}}}   }.
    \label{eq_cl_NO}
\end{equation}
For D1 studied in the main text, including the oxide would reduce $C'$ by $\sim 20\%$. Correspondingly, the $\sigma$ values extracted are overall smaller than those in Fig.~3 of the main text. Nevertheless, as shown in Fig.~\ref{Fig_Result_NbOxide}, the data still feature nodeless behavior and the parameters extracted from fitting are comparable to those in the main text. Last, we note that for future studies, the uncertainties from the oxide layer can be mitigated using thicker bottom hBN flakes.

\bibliography{reference}